\documentclass[review]{elsarticle}

\usepackage{hyperref}
\usepackage{amsmath}

\journal{Technical Physics Journal}

\begin{document}

\begin{frontmatter}

\title{ An Algorithm  for the calculation of non-isotropic collision integral matrix elements of the non-linear Boltzmann equation by the use of recurrence relations.}


\author[aff1]{I.A.Ender}
\author[aff2]{L.A.Bakaleinikov}
\author[aff2]{E.Yu.Flegontova}
\author[aff2]{A.B.Gerasimenko\corref{corrauth}}
\cortext[corrauth]{Corresponding author}
\ead{gerasimenko.alexander@mail.ioffe.ru}

\address[aff1]{Saint-Petersburg State University, Saint-Petersburg, Russia}
\address[aff2]{Ioffe Institute, Saint-Petersburg, Russia}

\begin{abstract}
 An algorithm for sequential calculation of non-isotropic matrix elements of the collision integral which are necessary for the solution of the non-linear Boltzmann equation by moment method  is proposed. Isotropic matrix elements that we believe are known, are starting  ones. The procedure is valid  for any  interaction law and any mass ratio of the colliding particles.
\end{abstract}


\end{frontmatter}


\section{Introduction}

Many of today's applications and technical issues require in-depth study of kinetic processes in gas mixtures. Kinetic approach is required in problems where DF strongly deviates from Maxwellian or is anisotropic. These problems include the kinetic description of the structure of shock waves and non-stationary processes of their interaction. One more problem of the same kind is to describe the time-dependent transport processes in low-temperature plasma.

One of the most effective methods of calculating the distribution function is a moment method. It is based on  DF expansion  in a set of  basis functions. The Boltzmann equation is thus reduced to a system of equations for the expansion coefficients. The  Boltzmann collision integral is replaced by a matrix whose elements are  coefficients of the collision integral expansion in the basis functions.

Development of the moment method is primarily due to Barnett \cite{Baa35, Ba35}.  It was in his works where non-linear system of moment equations was obtained first. Products of spherical harmonics by Sonine polynomials have been selected as basis functions. These are orthogonal with respect to the Maxwellian weight function.  Later, this set of basis functions  became known as Barnett functions. 
 Non-linear matrix elements of the collision integral (MEs) were considered in his work in this basis. 
As Barnett noted formula  obtained for the non-linear MEs calculation  is extremely cumbersome. 
For this reason, Barnett and his followers (see., Eg, \cite{R0PWM71}) in concrete calculations were limited to models of Maxwell molecules and hard spheres with $ l \leq 2 $, where $ l $  is the order of Legendre polynomial.
 
In 1966, Kumar \cite{Ku66} analysed various systems of polynomials that are used in the DF expansion in the kinetic theory of gases. He showed that the most cost-effective is the system of Barnett functions. Kumar also advanced in the study of the structure of the non-linear collision integral. He proposed to use the Talmi transformation (which had previously been used successfully in the quantum theory) for the calculation of non-linear MEs.

Kumar's ideas were further developed in the work of scientists of the Australian school \cite{RN, Ness, WNR, LRW}. In these studies the moment method is mainly used to obtain transport coefficients  describing the motion of charged particles in external fields. Despite the fact that in this case we consider a linear collision integral, until recently, it was not possible to calculate a sufficient number of terms of the DF expansion.

The problem of calculation of MEs with large indices is still significant. So in the relatively late work by 
Shizgal et al. \cite{SD} the calculation of linear isotropic MEs was considered. The authors suggested an algorithm that allows to calculate isotropic linear MEs with indices $ r \leq 50$ where $ r $ is the Sonine polinomial order of the DF expansion.  

In contrast to papers mentioned above,  where only several dozens of moments were calculated, we introduce the method for sequential calculation of MEs. This method allows to obtain dozens of thousands of matrix elements  \cite{Phf, book}. We used Burnett functions as basis functions. The set of basis functions is determined by the choice of temperature and mean velocity of weight Maxwellian. Recurrence relations for MEs were obtained by the use of fundamental principle of the collision integral invariance with respect to the choice of basis. 
These relations could be divided into two groups: temperature (from the invariance with respect to temperature choice) and velocity (from the invariance with respect to the choice of mean velocity value).

Velocity relations are algebraic, while temperature relations in the general case include the derivative with respect to the temperature. In the case of power interaction potentials all relations between MEs (both velocity and temperature) become algebraic. Recurrence procedure for sequential MEs calculation was developed for the case of corresponding interaction cross sections. In such a way all non-linear MEs  (both  isotropic, corresponding to the velocity isotropic DF, and non-isotropic) could be found if linear isotropic MEs are known.

Implementation of this approach let us calculate MEs with almost as high as desired indices. We have tabulated MEs for a number of power and quasi-power interaction laws. The result is a fast and accurate calculation of highly non-equilibrium distribution function in a variety of problems such as non-linear relaxation (DF was calculated up to 10-20 thermal velocities \cite{Phf}), calculation of transport coefficients and DF in the case of strong constant and periodic external fields. We would like to note that in the latter case  significant advance was achieved due to the transition to modified moment method. In this method temperature of weight Maxwellian of charged particles is different from temperature of background gas and changes with time \cite{JTF2016}. In this case we need special subset of nonlinear MEs from single-temperature moment method to calculate linear MEs corresponding to new temperature basis.

It should be noted that two-temperature moment method is widely applied to linear problems 
\cite{WNR, Mas2, viehland2012}. 
At the same time, sets of MEs used in these works do not provide 
accurate DF calculation in high energy region and at high fields. 

We emphasize that problems discussed in our previous works were solved for model cross sections  that correspond to power  or quasy-power interaction laws. The purpose of this article is generalisation of MEs calculation procedure for arbitrary interaction potentials and arbitrary mass ratios of interacting particles.

It will be shown that all non-isotropic MEs could be found with the use of velocity recurrence relations if all isotropic MEs (linear and non-linear) are known.

In the case of arbitrary interaction cross sections, temperature  recurrence relations include temperature derivations and it is convenient to expand isotropic MEs in $\Omega$-integrals known from kinetic theory \cite{ChC, FK} to transform relations to algebraic form. In such a way coefficients of MEs expansions turn out to be related with recurrence relations but not MEs itself. Isotropic MEs found by this method are starting for the velocity recurrence procedure which does not depend on interaction potential choice. As a result the recurrence procedure  is divided in two parts: calculation of starting MEs (linear and non-linear) in the form of finite sum of $\Omega$- integrals and calculation of all remaining MEs with the use of algebraic velocity recurrence relations. 
Note that in the second part of the recurrence procedure we deal with MEs only and not with $\Omega$-integrals expansion coefficients. It reduces the number of indexes  significantly.

In this article the recurrence procedure for successive calculation of non-isotropic MEs will be demonstrated. Starting isotropic MEs we will assume known.  The next article will be focused on their calculation. They will be represented as finite sums of $\Omega$-integrals.

\section{Basic relations}

Let us observe the mixture of two sorts of particles  $a$ and $b$ with masses  $m_a$ and $m_b$. In this case non-linear collision integrals take the form
\begin{equation}
\hat{I}(f_a,f_b)=n_a({\bf r}, t)n_b({\bf r}, t)\int \left(f_a({\bf v}_1)f_b({\bf
v}_2) - f_a({\bf v})
f_b({\bf v}') \right) g\sigma_{ab} (g,\theta)\ d{\bf v}'\ d{\bf k}.
\label{E1aP2.01}
\end{equation}
Here $ f_a $, $f_b $ are velocity distribution functions of particles of the sorts $a$ and $b$,
$n_a$ and  $n_b$ are particle number densities. Distribution functions are normalized to unity.

Particles velocities before and after collisions are related as
\begin{displaymath}
{\bf v}_1={\bf v}_0 -\mu_b{\bf  k}g,\qquad    {\bf  v}_2={\bf  v}_0+\mu_a{\bf
k}g,\qquad   {\bf v}_0=(\mu_a {\bf v}+{\mu_b \bf v}'),
\end{displaymath}
\begin{equation}
\mu_a=(m_a)/(m_a+m_b), \qquad   \mu_b=(m_b)/(m_a+m_b) ,   
\label{eqmass}
\end{equation}
\begin{equation}
{\bf g} = {\bf v}_1 - {\bf v}_2, \quad {\bf  g}'={\bf  v}-{\bf  v}', \quad {g} = {g}' ,
\nonumber
\end{equation}
where $\mu_a={m_a}/(m_a+m_b)$, $\mu_b={m_b}/(m_a+m_b)$, 
{\bf k} is unity vector, directed along ${\bf g}$. 
Scattering angle $\theta$ is given by  $\cos\theta={\bf
k}  \cdot  {\bf   g'}/g$,  and   $\sigma_{ab} (g,\theta)$  is differential cross section.

Collision integral can be written as
\begin{equation}
\hat{I}^a_{a,b}(f_a,f_b)=n_a n_b \int \int G^a_{a,b}\left({\bf v},{\bf v}_{1} , {\bf
v}_{2} \right)f_a({\bf v}_{1} )f_b({\bf v}_{2} )d{\bf v}_{1} d{\bf v}_{2} .   
\label{bakeq2}
\end{equation}
Here $G^a_{a,b}({\bf v}, {\bf v}_1, {\bf v}_2)$ is the kernel of collision integral depending on velocity vectors \cite{PP2015}. This kernel  contains delta-function which  provides compliance with the  energy conservation law.
We represent the distribution function in the form of an expansion in spherical harmonics
\begin{equation}
f_{a,b}\left({\bf v}\right)=\sum _{l=0}^{\infty }\sum _{m=0}^{l}\sum
_{i=0}^{1}(f_{a,b})_{l,m}^{i} \left(v\right)
Y_{l,m}^{i} \left(\Theta ,\varphi \right) , 
\label{bakeq3}
\end{equation}
$$
Y_{lm}^{0} (\Theta ,\varphi )=P_{l}^{m} (\cos \Theta )\cos m\varphi , \quad
Y_{lm}^{1} (\Theta ,\varphi )=P_{l}^{m} (\cos \Theta )\sin m\varphi .
$$
Here $ P^m_l (\cos {\Theta}) $  are associated Legendre polynomials, $ \Theta $,
$ \varphi $  are angular coordinates  of the vector $ {\bf v} $.
When using the expansion (\ref{bakeq3}) collision integral
takes the form 
\begin{eqnarray}
\hat{I}(f_a,f_b) =
\nonumber \\
= n_a({\bf r}, t) n_b({\bf r}, t) \int _{0}^{\infty }\int _{0}^{\infty }\left(\sum
_{l_{1} ,l_{2} ,m_{1} ,m_{2} ,i_{1} ,i_{2} }(G^a_{a,b})_{l_{1} ,m_{1} ,i_{1} ,l_{2}
,m_{2} ,i_{2} }^{l,m,i}  \left(v,v_{1} ,v_{2} \right) \right. \times \nonumber \\
\left. \times  (f_a)_{l_{1} ,m_{1} }^{i_{1} } \left(v_1\right)(f_b)_{l_{2} ,m_{2}
}^{i_{2} } \left(v_2\right) \right)
 v_{1}^{2} v_{2}^{2} dv_{1} dv_{2} .
\label{bakeq4}
\end{eqnarray}

As can be seen from (\ref{bakeq4}), a five-time collision integral is replaced by a sum of simpler two-time integrated operators. Kernels of these operators $ (G^a_{a, b})_{l_{1}, m_{1}, i_{1}, l_{2}, M_{2}, i_{2}}^{l, m, i} $, depending only on  velocity modules $v$, $ v_1 $, $ v_2 $, are the projections of kernel, which depends on velocity  vectors, on spherical harmonics. If we further expand $ (f_{a, b})_{l, m}^{i} \left (v \right) $ in the basis functions, 
depending on the  velocity module, the collision integral is replaced by the numerical matrix. 
We use Sonine (Laguerre) polinomials $ S_{l + 1/2}^{r} (x^2) $ as basis functions. Thus DF takes the form. 
\begin{eqnarray}
f_a({\bf c}_a,{\bf r},t)=M_a\sum C^{i,a}_{rlm}({\bf r},t)H^{i}_{rlm}
({\bf c}_a), \quad
{\bf c}_a=\sqrt{\frac{m_a}{2kT}} ({\bf v}-{\bf u}),
\label{q1E1}
\end{eqnarray}
\[
M_a=\left(\frac{m_a}{2kT\pi}\right)^{3/2}e^{-c_a^2},\]
\begin{equation}
H_j({\bf c}_a)=Y_{lm}^i(\Theta ,\varphi )c_a^lS^{r}_{l+1/2}(c_a^2), \quad i=0,1; 
\label{q1E12}
\end{equation}
%
%
Here the index $ j $ is composed of four indices ($ r, l, m, i $), $ M_a $ is weight Maxwellian with temperature $ T $ and 
 mean velocity {\bf u}. Similarly, the expansion coefficient of DF of $ b $ th component  $ f_b $ is written.

The Boltzmann equation in this case is replaced by an infinite system of moment equations. For $ a $ th 
component of the mixture DF expansion coefficients $ C_j ^ a $ are 
\begin{equation}
\frac{D_M(C^a_j)}{Dt}=\sum_{j_1,j_2}(K^a_{a,b})^j_{j_1,j_2}C^a_{j_1}C^b_{j_2}+\sum_{j_1,j_2}(K^a_{a,a})^j_{j_1,j_2}C^a_{j_1}C^a_{j_2}.
\label{q1e3}
\end{equation}
The explicit form of the differential operator $ D_M (C^a_j) / Dt $ can be found in \cite{ender2009}. Non-linear matrix elements of the collision integral $ (K^a_{a, b})_{j_1, j_2}^{j} $ are defined as follows:
\begin{eqnarray}
(K^a_{a,b})_{j_1,j_2}^{j} &=& \int H_j({\bf c}_a) \hat {I}(M_a H_{j_1}({\bf c}_a),M_b
H_{j_2}({\bf c}_b))\ d^3v/g_j,\nonumber \\
g_j &=& \int \pi^{-3/2} e^{-c^2}H_j^2({\bf c}) \ d^3c .
\label{q1E4}
\end{eqnarray}
They are related with kernels by 
\begin{eqnarray}
(K^a_{a,b})_{j_1,j_2}^{j} = \int \int \int 
(G^a_{a,b})^{l,m,i}_{l_1,m_1,i_1,l_2,m_2,i_2}(v,v_1,v_2) \times
\nonumber \\
\times S_{l+1/2}^{r}(c^2) M(c_1) S_{l_1+1/2}^{r}(c_1^2) M(c_2) S_{l_1+1/2}^{r}(c_2^2)
c^2 c_1^2 c_2^2  dc dc_1 dc_2 / \sigma_{rl} ,
\label{q1E4a}
\end{eqnarray}
where $M(c)= \pi^{-3/2} e^{-c^2}$.
Note the fact (see \cite{book}), that for 
any particles undirected in space, 
i.e. in the case of the scattering cross-sections depending on the two variables only (relative velocity module and scattering angle) arbitrary 
MEs are proportional to the corresponding axially symmetric MEs
(ie, MEs with indices $ m = m_1 = m_2 = 0 $):
\begin{equation}
 (K^a_{a,b})^{r,l,m,i}_{r_1,l_1,m_1,i_1,r_2,l_2,m_2,i_2}= 
\check Z^{lmi}_{l_1m_1i_1,l_2m_2i_2}(K^a_{a,b})^{r,l}_{r_1,l_1,r_2,l_2}.
\label{dsh5}
\end{equation}
Here and below we omit indices $ m $, $ i $ in the axisymmetric MEs notation. 
The essential is that numbers $ \check Z ^ {lmi} _ {l_1m_1i_1, l_2m_2i_2} $ can differ from zero only if
\begin{equation}
|l_l-l_2|\le l \le l_1+l_2, \qquad (-1)^{l+l_1+l_2}=1.
\label{OTG}
\end{equation}
In addition, conditions $ m = | m_1 \pm m_2 | $, $ (- 1)^{i + i_1 + i_2} = 1 $, $ m \leq l $, $ m_1 \leq l_1 $, $ m_2 \leq l_2 $ must be satisfied. These conditions, together with (\ref{OTG}) constitute the generalized Hecke theorem (GHT) \cite{ender2003}, which expands Hecke theorem \cite{He17, He22} to the non-linear case. For linear ME when $ j_1 $ or $ j_2 $ are equal to zero, $ \check Z = $ 1 and, respectively, 
either $ (l_2, m_2, i_2) = (l, m, i) $, or $ (l_1, m_1, i_1) = (l, m, i) $.

When DF is independent of velocity direction, only the matrix elements of the form $ K^{r, 0}_{r_{1}, 0, r_{2}, 0} $ 
 are non-zero. Then we will call them isotropic. Matrix elements with at least one non-zero index $ l $, $ l_1 $ or $ l_2 $
we will call non-isotropic.

Linear MEs correspond to the linear collision integral and may be  of the first ($ (K^a_{a, b})_{j, 0}^{i} $) 
 or second ($ (K^a_{a, b})_{ 0, k}^{i} $) kind.
Let us denote them  $ (\Lambda^a_{a, b}) $. In the axially symmetric case we have
\begin{equation}
(\Lambda^a_{a,b})_{r,r_1,l}^{(1)}=(K^a_{a,b})_{r_1,l,0,0}^{r,l};\quad
(\Lambda^a_{a,b})_{r,r_2,l}^{(2)}=(K^a_{a,b})_{0,0,r_2,l}^{r,l}.
\label{q1E5}
\end{equation}

Using (\ref{dsh5}) and (\ref{rme4}), is easy to show that kernels of the collision integral obtained 
by projecting $ G^a_{a, b} \left({\bf v}, {\bf v }_{1}, {\bf v}_{2} \right) $ on spherical harmonics, are linear combinations of axially symmetric kernels $ (G^a_{a, b})^l_{l_1, l_2} (v, v_1, v_2) $
\begin{equation}
 (G^a_{a,b})^{l,m,i}_{l_1,m_1,i_1,l_2,m_2,i_2}(v,v_1,v_2)= 
\check Z^{m,i}_{m_1,i_1,m_2,i_2}(l,l_1,l_2)(G^a_{a,b})^{l}_{l_1,l_2}(v,v_1,v_2),
\end{equation}
Kernels $ (G^a_{a, b})_{l_{1}, l_{2}}^{l} $ are defined as coefficients of $ G^a_{a, b} \left ({\bf v }, {\bf v}_{1}, {\bf v}_{2} \right) $ expansion in Legendre polynomials
\begin{eqnarray}
(G^a_{a,b})_{l_{1} ,l_{2} }^{l} \left(v,v_{1} ,v_{2} \right)=4\pi \int _{-1}^{1}\int
_{-1}^{1}\int _{-1}^{1}\frac{P_{l} (\cos \Theta)}
{\left\| P_{l} \right\| ^{2} }    G^a_{a,b}\left({\bf v},{\bf v}_{1} ,{\bf v}_{2}
\right)  \times \nonumber \\
\times P_{l_{1} } (\cos \Theta _{1})  P_{l_{2}} (\cos \Theta _{2}) 
d\cos \Theta d\cos \Theta _{1} d\cos \Theta _{2} .
\label{bakeq5}
\end{eqnarray}
It follows from the written above, that kernels (\ref{bakeq5}) could be represented as
\begin{eqnarray}
&&(G^a_{a,b})^l_{l_1,l_2}  \left(v,v_{1} ,v_{2} \right) = \nonumber \\
&&M(c)\sum\limits_{r,r_1,r_2
}c^l S_{l + 1 / 2}^r (c^2)
(K^a_{a,b})^{r,l}_{r_1,l_1,r_2,l_2}
\frac{c_1^{l_1} S_{l_1 + 1 / 2}^{r_l } (c_1 ^2)}{\sigma _{r_1 l_1 }} 
\frac{c_2^{l_2} S_{l_2 + 1 / 2}^{r_2 } (c_2 ^2)}{\sigma _{r_2 l_2 }} , \quad
\label{rme4}
\end{eqnarray}
where $ \sigma_{r l} $  are normalization factors  of Sonine polynomials.
 
In \cite{book, ender2007} from the invariance of the collision integral with respect to the choice of temperature and 
the mean velocity module of weight Maxwellian two groups of  recurrence relations were obtained
The first one are temperature relations 
\begin{eqnarray}
&&( T\frac{d}{d T} -R) (K^a_{a,b})^{r,l}_{r_1,l_1,r_2,l_2}= r
(K^a_{a,b})^{r-1,l}_{r_1,l_1,r_2,l_2}-
\nonumber\\
&&-(r_1+1)(K^a_{a,b})^{r,l}_{r_1+1,l_1,r_2,l_2}-
(r_2+1)(K^a_{a,b})^{r,l}_{r_1,l_1,r_2+1,l_2},
\label{q1E1P4.11}
\end{eqnarray}%
\begin{eqnarray} 
R=r_{1}+r_{2}-r+(l_{1}+l_{2}-l)/2 ,
\nonumber
\end{eqnarray}
 and the second one are velocity relations
\begin{eqnarray}
\beta(l-1)(K^a_{a,b})^{r ,l
-1}_{r_{1},l_{1},r_{2},l_{2}}+\gamma(r-1,l+1)(K^a_{a,b})^{r -1,l
+1}_{r_{1},l_1,r_{2},l_{2}}- \nonumber \\
-\beta(l_{1})(K^a_{a,b})^{r,l}_{r_{1},l_{1}+1,r_{2},l_{2}}-
-\gamma(r_{1},l_{1})(K^a_{a,b})^{r ,l}_{r_{1}+1,l_{1}-1,r_{2},l_{2}} - \nonumber \\
-\sqrt{\frac {m_b}{m_a}} \left(\beta(l_{2})(K^a_{a,b})^{r ,l}_{r_{1}, l_{1},r_{2},l_{2}+1}+
\gamma(r_{2},l_{2})(K^a_{a,b})^{r ,l}_{r_{1},l_{1},r_{2}+1,l_{2}- 1}\right)=0,\nonumber \\
\beta(l)=-\frac{l+1}{2l+1}, \quad \quad\quad\gamma(r,l)=\frac{(r+1)l}{2l+1}.
\label{q1E1P4.12}
\end{eqnarray}
In work \cite{Phf, book} on the basis of (\ref{q1E1P4.11}), (\ref{q1E1P4.12}) the authors developed a 
recurrence procedure using a simple analytical expression for linear isotropic MEs for power 
interaction laws. For arbitrary interaction potentials we offer a modified 
 recurrence procedure. The calculation is divided into two stages. In the first stage temperature 
relations are used to construct linear and non-linear isotropic MEs via expansion in $ \Omega $ -integrals 
with the recurrence relations for MEs reduced to the algebraic relations for the expansion coefficients.
 The most efficient is the use of expansion coefficients of linear isotropic MEs 
of the second kind $ (\Lambda^a_{a, b})_{r, r_2, l} ^ {(2)} $ as the 
starting. At the second  stage calculated isotropic MEs are used as starting to build non-isotropic MEs with non-zero indices $ l $, $ l_1 $, $ l_2 $. Thus, a specific type of interaction cross section is used only at the stage 
{isotropic MEs calculation. }
This article will describe the second universal part of the algorithm. Our next article will be focused on  the calculation of starting isotropic MEs.

\section{ Calculation of non-isotropic matrix elements } 

We show that  velocity relation (\ref{q1E1P4.12}) allow us to consistently express MEs with arbitrary indices directly through isotropic MEs without recourse to the $ \Omega $ -integrals at each step of the second stage of the 
 recurrence procedure. For brevity, in the MEs notation we shall omit the indices $ a $, $ b $, indicating the 
 sort of interacting particles.

We introduce, as in \cite{book}, the parameters $ R, p, q, \lambda, \tau, \nu $:
\begin{displaymath}
R=r_1 -r + \left( l_{1}+l_{2}-l \right)/2, \qquad p=r_{1}, \qquad q=r_{2},
\end{displaymath} 
\begin{equation}
\lambda = \left( l_{1}+l_{2}+l \right)/2,  \qquad \tau=\left( l-l_{1}+l_{2}
\right)/2,  \qquad \nu = l_{2}.
\label{E1P4.9}
\end{equation}
It is easy to express the original 
 indices through these 
 parameters:
\begin{displaymath}
r=p-R-\tau+\nu, \qquad r_{1}=p, \qquad r_{2}=q, 
\end{displaymath} 
\begin{equation}
l=\lambda-\nu+\tau \qquad l_{1}=\lambda-\tau, \qquad l_{2}=\nu.
\label{E1P4.10}
\end{equation}

The non-negativity of $ l $, $ l_1 $, $ l_2 $ implies that $ \lambda \geq 0 $, $ \nu \geq 0 $, and 
it follows from the evenness of the sum $ l + l_1 + l_2 $ (see (\ref{OTG})) 
that $ \lambda $, $ \nu $ 
 are integers. It can be shown that other 
 GHT conditions (\ref{OTG}) are equivalent to inequalities
\begin{equation}
0 \leq \tau \leq \nu \leq \lambda.  
\label{eqchain}
\end{equation}
 Let us replace in the 
(\ref{q1E1P4.12}) $ l_2 $ by $ l_2 - 1$, and rearrange the terms so as to express the $ K^{r, l}_{r_{1}, l_{1}, r_{2}, l_{2}} $ 
 through the rest of MEs, and then rewrite (\ref{q1E1P4.12}), using parameters defined above:
\begin{eqnarray}
&&\sqrt{\frac{m_b}{m_a}}\beta(\nu-1)K^{p-R-\tau+\nu,\lambda-\nu+\tau}_{p,\lambda-\tau,q,\nu}= \nonumber \\
&& \gamma(p-R-\tau+\nu-1,\lambda-\nu+\tau+1)K^{p-R-\tau+\nu-1,\lambda-\nu+\tau+1}_{p,\lambda-\tau,q,\nu-1}- \nonumber \\
&&-\beta(\lambda-\tau)K^{p-R-\tau+\nu ,\lambda-\nu+\tau}_{p,\lambda-\tau+1,q,\nu-1}
+\left\lbrace \beta(\lambda-\nu+\tau-1)K^{p-R-\tau+\nu,\lambda-\nu+\tau
-1}_{p,\lambda-\tau,q,\nu-1}- \right. \nonumber \\
&&\left. -\gamma(p,\lambda-\tau)K^{p-R-\tau+\nu,\lambda-\nu+\tau}_{p+1,\lambda-\tau-1,q,\nu-1} - \sqrt{\frac{m_b}{m_a}}\gamma(q,\nu-1)K^{p-R-\tau+\nu,\lambda-\nu+\tau}_{p,\lambda-\tau,q+1,\nu-2}
\right\rbrace. 
\label{E1P4.11}
\end{eqnarray}

Note that in braces are MEs, for which 
the sum of the indices corresponding to the expansion in Legendre polynomials is 
$ 2 \lambda-2$, that is 
 less by 2 than the sum of indices for MEs outside of braces equal to $\lambda$. 
 Below MEs and kernels with sum of indices $l+l_1+l_2=\lambda$ will be referred to as belonging to the layer $\lambda$.
We denote the expression in braces $ Q^R_{p, \lambda, q} (\tau, \nu) $. We assume that it is found in the previous step of the 
recurrence procedure. 
In the first step in braces are isotropic MEs, which are starting. We fix the parameters $ \lambda $, $ p $, $ q $ and $ R \leq p $. Let $ \tau = 0 $ and write 
(\ref{E1P4.11}) for $ \nu \geq 1 $ (with $ \nu = 0 $ left and right side of the equation are equal to zero)
\begin{eqnarray}
\sqrt{\frac{m_b}{m_a}}\beta(\nu-1)K^{p-R+\nu,\lambda-\nu}_{p,\lambda,q,\nu}&=&
\gamma(p-R+\nu-1,\lambda-\nu+1)K^{p-R+\nu-1,\lambda-\nu+1}_{p,\lambda,q,\nu-1}- \nonumber \\
&&-\beta(\lambda)K^{p-R+\nu
,\lambda-\nu}_{p,\lambda+1,q,\nu-1}+Q^R_{p,\lambda,q}(0,\nu)
\label{E1P4.12}
\end{eqnarray}

The second term on the right in this equation is equal to zero, since by virtue of (\ref{OTG}) MEs with indices that do not satisfy the inequality $ \vert l_{1} -l_{2} \vert \leq l $, vanish. Using mathematical induction method, it is easy to show that for $ 1 \leq \nu \leq \lambda $ equality (\ref{E1P4.12}) can be rewritten as
\begin{equation}
K^{p-R+\nu,\lambda-\nu}_{p,\lambda,q,\nu}=
L^{R}_{p,\lambda,q}(0,\nu)K^{p-R,\lambda}_{p,\lambda,q,0}+M^{R}_{p,\lambda,q}(0,\nu),
\label{E1P4.13}
\end{equation}
where
\begin{equation}
L^{R}_{p,\lambda,q}(0,\nu)=\frac{\gamma(p-R+\nu-1,\lambda-\nu+1)}{\beta(\nu-1)\sqrt{m_b/m_a}}L^{R}_{p,\lambda,q}(0,\nu-1),
\end{equation}
\begin{eqnarray}
M^{R}_{p,\lambda,q}(0,\nu)&=&\frac{\gamma(p-R+\nu-1,\lambda-\nu+1)}{\beta(\nu-1)\sqrt{m_b/m_a}}M^{R}_{p,\lambda,q}(0,\nu-1)+ \nonumber \\
&+&\frac{Q^R_{p,\lambda,q}(0,\nu)}{\beta(\nu-1)\sqrt{m_b/m_a}}.
\label{E1P4.14}
\end{eqnarray}
That is, all MEs 
$ K^{p-R + \nu, \lambda- \nu}_{p, \lambda, q, \nu} $ are expressed through the $ K^{p-R, \lambda}_{p, \lambda, q, 0} $ and known functions $ Q^R_{p, \lambda, q} (0, \nu) $. Let us now set  in (\ref{E1P4.11}) $ \tau = 1 $:
\begin{eqnarray}
&&\beta(\nu-1)\sqrt{m_b/m_a}K^{p-R-1+\nu,\lambda-\nu+1}_{p,\lambda-1,q,\nu}=\nonumber \\
&&=\gamma(p-R+\nu-2,\lambda-\nu+2)K^{p-R+\nu-2,\lambda-\nu+2}_{p,\lambda-1,q,\nu-1}- \nonumber \\
&&-\beta(\lambda-1)K^{p-R-1+\nu,\lambda-\nu+1}_{p,\lambda,q,\nu-1}+Q^R_{p,\lambda,q}(1,\nu).
\label{E1P4.15}
\end{eqnarray}
When $ \nu = 1 $ we have
\begin{eqnarray}
\beta(0)\sqrt{m_b/m_a}K^{p-R,\lambda}_{p,\lambda-1,q,1}&=&
\gamma(p-R-1,\lambda+1)K^{p-R-1,\lambda+1}_{p,\lambda-1,q,0}- \nonumber \\
&-&\beta(\lambda-1)K^{p-R ,\lambda}_{p,\lambda,q,0}+Q^R_{p,\lambda,q}(1,1).
\label{E1P4.16}
\end{eqnarray}
The first term on the right hand side vanishes, since matrix element indices do nor satisfy the condition $ l \leq l_{1} + l_{2} $ in (\ref{OTG}). 
Thus, $ K^{p-R, \lambda} _ {p, \lambda-1, q, 1} $ is represented as
\begin{equation}
K^{p-R,\lambda}_{p,\lambda-1,q,1}
=L^{R}_{p,\lambda,q}(1,1)K^{p-R ,\lambda}_{p,\lambda,q,0}+M^R_{p,\lambda,q}(1,1) ,
\label{E1P4.17}
\end{equation} 
where
\begin{equation}
L^{R}_{p,\lambda,q}(1,1)=-\frac{\beta(\lambda-1)}{\beta(0)\sqrt{m_b/m_a}}, \qquad
M^R_{p,\lambda,q}(1,1)=\frac{Q^R_{p,\lambda,q}(1,1)}{\beta(0)\sqrt{m_b/m_a}}. 
\label{E1P4.18}
\end{equation}
Consistently increasing $ \nu $ in the range $ 1 <\nu \leq \lambda $ and using induction method  it is easy to show that all MEs of the type $ K^{p-R + \nu-1, \lambda - (\nu-1)} _ {p, \lambda-1, q, \nu} $ can also be expressed through the $ K^{p-R, \lambda} _ {p, \lambda, q, 0} $ and known functions $ Q^R_{p, \lambda, q} (1, \nu) $:
\begin{equation}
K^{p-R+\nu-1,\lambda-(\nu-1)}_{p,\lambda-1,q,\nu}=
L^{R}_{p,\lambda,q}(1,\nu)K^{p-R ,\lambda}_{p,\lambda,q,0}+M^R_{p,\lambda,q}(1,\nu),
\label{E1P4.20}
\end{equation}
where
\begin{eqnarray}
L^{R}_{p,\lambda,q}(1,\nu)&=&\frac{\gamma(p-R+\nu-2,\lambda-\nu+2)}{\beta(\nu-1)\sqrt{m_b/m_a}}L^R_{p,\lambda,q}(1,\nu-1)- \nonumber \\
&-&\frac{\beta(\lambda-1)}{\beta(\nu-1)\sqrt{m_b/m_a}}L^R_{p,\lambda,q}(0,\nu-1),
\label{eqLR}
\end{eqnarray}
\begin{displaymath}
M^R_{p,\lambda,q}(1,\nu)=\frac{\gamma(p-R+\nu-2,\lambda-\nu+2)}{\beta(\nu-1)\sqrt{m_b/m_a}}M^R_{p,\lambda,q}(1,\nu-1)-
\end{displaymath}
\begin{equation}
-\frac{\beta(\lambda-1)}{\beta(\nu-1)\sqrt{m_b/m_a}}M^R_{p,\lambda,q}(0,\nu-1)
+\frac{Q^R_{p,\lambda,q}(1,\nu)}{\beta(\nu-1)\sqrt{m_b/m_a}}.
\label{eqMR}
\end{equation}
When $ \nu = \lambda + 1 $ matrix element $ K^{p-R + \lambda, 0} _ {p, \lambda-1, q, \lambda + 1} $
on the left side of the recurrence relation, goes to zero, which gives an equation for determining the $ K^{p-R, \lambda} _ {p, \lambda, q, 0} $
\begin{displaymath}
L^{R}_{p,\lambda,q}(1,\lambda+1)K^{p-R
,\lambda}_{p,\lambda,q,0}+M^R_{p,\lambda,q}(1,\lambda+1)=0 .
\end{displaymath}
Here
\begin{displaymath}
K^{p-R,\lambda}_{p,\lambda,q,0}=-\frac{M^R_{p,\lambda,q}(1,\lambda+1)}{L^{R}_{p,\lambda,q}(1,\lambda+1)} .
\end{displaymath}
Now from relations (\ref{E1P4.13}), (\ref{E1P4.17}), (\ref{E1P4.20}) 
all MEs corresponding to $ \tau = 0,1 $ can be found.

To find MEs corresponding to other values of $ \tau $, we consider the basic relation with the $ \tau> 1 $. 
When $ \nu = \tau $ relation (\ref{E1P4.11}) transforms to
\begin{eqnarray}
\beta(\tau-1)\sqrt{\frac {m_b}{m_a}} K^{p-R,\lambda}_{p,\lambda-\tau,q,\tau}&=&
\gamma(p-R-1,\lambda+1)K^{p-R-1,\lambda+1}_{p,\lambda-\tau,q,\tau-1}- \nonumber \\
&-&\beta(\lambda-\tau)K^{p-R
,\lambda}_{p,\lambda-\tau+1,q,\tau-1}+Q^R_{p,\lambda,q}(\tau,\tau).
\end{eqnarray}
The first term on the right hand side vanishes by virtue of (\ref{OTG}) (the sum of matrix element lower indices is less than the  upper  index). Thus 
\begin{equation}
K^{p-R,\lambda}_{p,\lambda-\tau,q,\tau}=L^R_{p,\lambda,q}(\tau,\tau)K^{p-R
,\lambda}_{p,\lambda-\tau+1,q,\tau-1}+M^R_{p,\lambda,q}(\tau,\tau)
\end{equation}
where
\begin{displaymath}
L^R_{p,\lambda,q}(\tau,\tau)=-\frac{\beta(\lambda-\tau)}{\beta(\tau-1)\sqrt{m_b/m_a}},
 \qquad 
M^R_{p,\lambda,q}(\tau,\tau)=\frac{Q^R_{p,\lambda,q}(\tau,\tau)}{\beta(\tau-1)\sqrt{m_b/m_a}}.
\end{displaymath}
Changing $ \tau $ from 2 to $ \lambda $, consistently find MEs 
$ K^{p-R, \lambda} _ {p, \lambda- \tau, q, \tau} $, because $ K^{p-R, \lambda} _ {p, \lambda-1, q, 1} $ 
are known. We now use the basic relation (\ref{E1P4.11}) with $ \nu> \tau $. It has the form
\begin{displaymath}
K^{p-R-\tau+\nu,\lambda-\nu+\tau}_{p,\lambda-\tau,q,\nu}=
\frac{\gamma(p-R-\tau+\nu-1,\lambda-\nu+\tau+1)}{\beta(\nu-1)\sqrt{m_b/m_a}}
K^{p-R-\tau+\nu-1,\lambda-\nu+\tau+1}_{p,\lambda-\tau,q,\nu-1}-
\end{displaymath}
\begin{equation}
-\frac{\beta(\lambda-\tau)}{\beta(\nu-1)\sqrt{m_b/m_a}}
K^{p-R-\tau+\nu ,\lambda-\nu+\tau}_{p,\lambda-\tau+1,q,\nu-1}+
\frac{Q^R_{p,\lambda,q} (\tau, \nu)}{\beta(\nu-1)\sqrt{m_b/m_a}}.
\label{eqtaunu}
\end{equation}
 For $ \tau $ in the range from $ 2 $ to $ \lambda $, 
by increasing $ \nu $ from $ \tau +1 $ to $ \lambda $, we find all MEs of the form $ K ^ {p-R- \tau + \nu, \lambda- \nu + \tau}_{p, \lambda- \tau, q, \nu} $. 
 It is possible because the matrix element in the first term on the right hand side of (\ref{eqtaunu}) was  obtained for the previous $\nu$ value, and the matrix element in the second term was found for the previous $\tau$ value.

To find MEs with all combinations of the indices, it is necessary to perform the above procedure for all possible $ p $ and $ q $. Note that in the function $ Q^R_ {p, \lambda, q} (\tau, \nu) $ in (\ref{E1P4.11}) the matrix element with indices $ p + 1 $, $ q + 1 $ is included. 
This means that with an increase of $ \lambda $ by 1 the ranges of  $ p $ and $ q $ parameters are reduced by one. If you know the starting isotropic MEs with indices $ 0 \leq r_1 \leq N $, $ 0 \leq r_2 \leq N $, then for a fixed value of $ \lambda $, you can find the MEs with indices $ 0 \leq r_1 \leq N - \lambda $, $ 0 \leq r_2 \leq N - \lambda $.

It was assumed above that $ R \leq p $, and the matrix element $ K^{p-R, \lambda}_{p, \lambda, q, 0} $ 
 in (\ref{E1P4.13}), (\ref{E1P4.17}), (\ref{E1P4.20}) is different from zero.
This corresponds to the condition $ \nu - \tau \leq r $. To get MEs with indices $ 0 \leq r <\nu - \tau $ it is necessary to consider the $ R $ in the range $ p <R \leq p + (\nu - \tau) $. 
 In the analogy with the procedure described above, we will 
 consistently increase $ \tau $ from zero to $ \lambda $, and $ \nu $ from $ R-p + \tau $ to $ \lambda $ for fixed $ p $, $ q $ and $ R> p $.

Consider the relation (\ref{E1P4.12}), into which the basic relation (\ref{E1P4.11}) transforms when $ \tau = 0 $. 
When $ \nu = R-p $, the first term on the right side in (\ref{E1P4.12}) is zero, and
\begin{equation}
K^{0,\lambda-R+p}_{p,\lambda,q,R-p}=
\frac{Q^R_{p,\lambda,q}(0,R-p)}{\beta(R-p-1)\sqrt{m_b/m_a}}.
\label{E1P4.21}
\end{equation}
For $ R-p +1 \leq \nu \leq \lambda $ MEs of  the type $ K ^ {p-R + \nu, \lambda-\nu} _ {p, \lambda, q, \nu} $ 
are  sequentially calculated using relation (\ref{E1P4.12}). Note that 
for $ \nu = \lambda +1 $ right side of (\ref{E1P4.12}) must vanish.

Let us now consider the  basic relation 
for $ 1 \leq \tau \leq \lambda $. In the case 
of $ \nu = R-p + \tau $, relation (\ref{E1P4.11}) becomes (\ref{eqtaunu}), where the first term on the right is zero, and the second term is known. Therefore
\begin{eqnarray}
K^{0,\lambda-R+p}_{p,\lambda-\tau,q,R-p+\tau}&=&
-\frac{\beta(\lambda-\tau)}{\beta(R-p+\tau-1)\sqrt{m_b/m_a}}
K^{0,\lambda-R+p}_{p,\lambda-\tau+1,q,R-p+\tau-1}+ \nonumber \\
&+&\frac{Q^R_{p,\lambda,q} (\tau, R-p+\tau)}{\beta(R-p+\tau-1)\sqrt{m_b/m_a}}.
\label{Rptaunu}
\end{eqnarray}
If $ R-p + \tau \leq \nu \leq \lambda $ MEs 
are found sequentially from the relation (\ref{eqtaunu}). When $ \nu = \lambda + 1 $ MEs should be zero.

Thus, all MEs with indices corresponding to the layer $ \lambda $ can be found from the 
recurrence procedure based on velocity relations.
Note that at the stage of a non-isotropic MEs calculation temperature relations (\ref{q1E1P4.11}) are not used. 
In \cite{book, ender2009} 
a recurrence procedure for calculating 
MEs in the case
of power interaction law is described.
In contrast to the method reported above it includes the use of the temperature relations (\ref{q1E1P4.11}) 
while calculating non-isotropic MEs. 
This is due to the fact that in the case of power potentials relations (\ref{q1E1P4.11}) turn out to be algebraic. 
For hard spheres model the comparison of MEs calculated by use of two methods 
(described here and reported in \cite{book, ender2009}) has been made and the results has coincided. 

\section{Temperature recurrence relations}

We emphasize once again that in the presently described procedure temperature recurrence relations were not used directly. Thus, they are additional relation between constructed MEs. We will prove that (\ref{q1E1P4.11}) are satisfied identically, and both procedures (as described in the \cite{book, ender2009} for power potential and proposed in this article) are equivalent. To this end, we show that if the starting isotropic MEs satisfy temperature relations (\ref{q1E1P4.11}), then all MEs satisfy these relations.

Note that the matrix elements obtained at fixed $ \lambda $ allow us to determine the kernels of the collision integral using (\ref{rme4}). 
In this case (\ref{q1E1P4.11}) is equivalent to
\begin{equation}
\left(2T \frac {\partial} {\partial T} - c \frac {\partial} {\partial c} - 
c_1 \frac {\partial} {\partial c_1} - c_2 \frac {\partial} {\partial c_2} \right)
G^l_{l_1,l_2} = 0.
\label{trk}
\end{equation}  
To show this, {it is} sufficient to substitute (\ref{rme4}) in (\ref{trk}) and  to
use {relations between} Laguerre polynomials $ L ^ {\alpha} _r (x) $ \cite{Ryzhik}
\begin{eqnarray}
x \frac {dL^{\alpha}_r (x)} {dx} &=& n L^{\alpha}_r (x) - (n + \alpha)
L^{\alpha}_{r-1} (x) = \nonumber \\
&=&(n+1) L^{\alpha}_{r+1} (x) - (n + \alpha + 1 - x) L^{\alpha}_r (x).
\end{eqnarray}
Laguerre polynomials $ L ^ {\alpha} _r (x) $ coincide with 
Sonine {polynomials} $ S ^ {r} _ {l + 1/2} (x) $ at $ \alpha = l + 1/2 $, where $ l $ 
{is integer.}
Since isotropic MEs satisfy (\ref{q1E1P4.11}), then the kernel $ G ^ 0_ {0,0} $ satisfies (\ref{trk}). We show that if (\ref{trk}) holds for kernels  in the layer $ \lambda - $ 1, then it is true for kernels  in the layer $ \lambda $, and hence corresponding matrix elements satisfy temperature relations.

As shown in \cite{DAN}, there are relations that link kernels in the layer $ \lambda $ with kernels in the layer $ \lambda - 1$. 
In particular, for the kernel $ G ^ {\lambda} _ {\lambda,  0}$  we have \cite{RecPr}
\begin{eqnarray}
G_{\lambda ,0}^{\lambda } &=&\frac{(-1)^{\lambda +1} }{(\lambda +1)}
\left(\hat{B}_{\lambda -1}^{(2)} (c)\right)^{-1} ...\left(\hat{B}_{1}^{(2)}
(c)\right)^{-1} \left(\hat{B}_{0}^{(2)} (c)\right)^{-1} \left(\hat{B}_{\lambda
-1}^{(3)} (c_{1} )\right)^{-1} \times \nonumber \\
&\times &  \left(\hat{B}_{0}^{(3)} (c_{2} )...\hat{B}_{\lambda
-2}^{(3)} (c_{2} )\hat{B}_{\lambda -1}^{(3)} (c_{2} )\right) \times \nonumber \\
&\times & \left(\hat{B}_{0}^{(2)} (c)\tilde{M_{\lambda}}  (1,\lambda )+\hat{B}_{\lambda
-1}^{(3)} (c_{1} )\tilde{M_{\lambda} } (0,\lambda )+\hat{B}_{\lambda }^{(4)} (c_{2}
)G_{\lambda -1,\lambda -1}^{0} \right),
\label{GrindEQ__45_}
\end{eqnarray}
where the operators $ \hat{B}^{(i)}_{j} (c) $, $ \left (\hat{B}_{j}^{(i)} (c) \right)^{-1} $  are defined by expressions
\begin{eqnarray}
&&\hat{B}_{j}^{(i)} (c) f(c) = a(i,j) \left( \frac {\partial} {\partial c} + \frac
{b(i,j)} {c} \right) f(c), \nonumber \\
&&\left( \hat{B}_{j}^{(i)} (c) \right)^{-1} f(c) = \frac {a(i,j)} {c^{b(i,j)}}
\int_{0}^c t^{b(i,j)}  f(t) dt , \nonumber
\end{eqnarray}
$ a(1,j) = j/(2j-1)$, $ a(2,j)=(j+1)/(2j+3)$, $ a(3,j)=(j+1)/(2j+1)$, $
a(4,j)=j/(2j+1) $, 
$ b(1,j) = b(4,j) = j-1 $, $ b(2,j) = b(3,j) = j+2 $. 
Functions $ \tilde {M_{\lambda}} (0, \lambda) $, $ \tilde {M_{\lambda}} (1, \lambda) $ are found from the recurrence relations
\begin{eqnarray}
\tilde{M_{\lambda} } (0,1)&=&-(\hat{B}_{0}^{(3)} (c_{2} ))^{-1} \hat{B}_{\lambda
}^{(4)} (c_{1} )G_{\lambda -1,0}^{\lambda -1} , \nonumber \\
\tilde{M_{\lambda} } (0,\nu )&=&(-\hat{B}_{\nu -1}^{(3)} (c_{2} ))^{-1}
\left[\hat{B}_{\lambda -\nu }^{(2)} (c)\tilde{M_{\lambda} } (0,\nu
-1)+ \right.\nonumber \\
&+&\left.\hat{B}_{\lambda }^{(4)} (c_{1} )G_{\lambda -1,\nu -1}^{\lambda -\nu } \right], 
  \quad  \nu =2,3,...,\lambda , \nonumber \\
\tilde{M_{\lambda} } (1,1)&=&-(\hat{B}_{0}^{(3)} (c_{2} ))^{-1} \hat{B}_{\lambda
}^{(1)} (c)G_{\lambda -1,0}^{\lambda -1} , \nonumber \\
\tilde{M_{\lambda} } (1,\nu )&=&-(\hat{B}_{\nu -1}^{(3)} (c_{2} ))^{-1}
\left[\hat{B}_{\lambda -\nu +1}^{(2)} (c)\tilde{M_{\lambda} } (1,\nu
-1)\right. +\nonumber \\
&+&\left.\hat{B}_{\lambda -1}^{(3)} (c_{1} )\tilde{M_{\lambda} } (0,\nu -1)\right. +\left. \hat{B}_{\lambda -\nu +1}^{(2)} (c)G_{\lambda -1,\nu -1}^{\lambda -\nu }
\right. +\nonumber \\
&+&\left.\hat{B}_{\lambda -1}^{(4)} (c_{1} )G_{\lambda -2,\nu -1}^{\lambda -\nu +1}
+\hat{B}_{\nu -1}^{(3)} (c_{2} )G_{\lambda -1,\nu -2}^{\lambda -\nu +1} \right].
\label{GrindEQ__35_}
\end{eqnarray}
Notice, that
\begin{eqnarray}
c\frac{\partial }{\partial c} \hat{B}_{\nu }^{(i)} (c)=\hat{B}_{\nu }^{(i)}
(c)\left(c\frac{\partial }{\partial c} -1\right), \nonumber \\
c\frac{\partial }{\partial c} \left(\hat{B}_{\nu }^{(i)} (c)\right)^{-1}
=\left(\hat{B}_{\nu }^{(i)} (c)\right)^{-1} \left(c\frac{\partial }{\partial c} +1
\right).
\label{difc}
\end{eqnarray}

Therefore
\[
\left(2T\frac{\partial }{\partial T} -c\frac{\partial }{\partial c} -c_{1}
\frac{\partial }{\partial c_{1} } -c_{1} \frac{\partial }{\partial c_{1} }
\right)G_{\lambda ,0}^{\lambda } =
\] 
\[
=\frac{(-1)^{\lambda +1} }{(\lambda +1)} \left(\hat{B}_{\lambda -1}^{(2)}
(c)\right)^{-1} ...\left(\hat{B}_{1}^{(2)} (c)\right)^{-1} \left(\hat{B}_{0}^{(2)}
(c)\right)^{-1} \times
\]
\[\times\left(\hat{B}_{\lambda -1}^{(3)} (c_{1} )\right)^{-1}
\left(\hat{B}_{0}^{(3)} (c_{2} )...\hat{B}_{\lambda -2}^{(3)} (c_{2}
)\hat{B}_{\lambda -1}^{(3)} (c_{2} )\right)\times 
\] 
\begin{equation}
\times \left( 
\hat{B}_{0}^{(2)} (c) \left( 2T\frac{\partial }{\partial T} -c\frac{\partial
}{\partial c} -c_{1} \frac{\partial }{\partial c_{1} } -c_{2} \frac{\partial
}{\partial c_{2} } \right) \tilde{M_{\lambda} } (1,\lambda )+ 
\right.
\nonumber
\end{equation}
\begin{equation}
\left.
\hat{B}_{\lambda -1}^{(3)} (c_{1} )\left(2T\frac{\partial }{\partial T}
-c\frac{\partial }{\partial c} -c_{1} \frac{\partial }{\partial c_{1} } -c_{2}
\frac{\partial }{\partial c_{2} } \right)\tilde{M_{\lambda} } (0,\lambda )+
\right.  
\nonumber
\end{equation}
\begin{equation}
\left.
+\hat{B}_{\lambda }^{(4)} (c_{2} )\left(2T\frac{\partial }{\partial T}
-c\frac{\partial }{\partial c} -c_{1} \frac{\partial }{\partial c_{1} } -c_{2}
\frac{\partial }{\partial c_{2} } \right)G_{\lambda -1,\lambda -1}^{0} \right).
\label{transop}
\end{equation}
The last term in parentheses is zero since kernels in layer $ \lambda -1 $ satisfy the relation (\ref{trk}). The first two 
terms vanish by virtue of formulas (\ref{GrindEQ__35_}), (\ref{difc}). That is, (\ref{trk}) is true 
for $ G_{\lambda, 0}^{\lambda} $.

The validity of (\ref{trk}) for the 
kernels $ G_{\lambda, \nu}^{\lambda - \nu} $, $ G_{\lambda -1, \nu}^{\lambda - \nu +1 } $, $ G_{\lambda - \tau, \tau} ^ {\lambda} $, $ G_{\lambda - \tau, \nu} ^ {\lambda + \tau - \nu} $ can be proved similarly, using formulas from \cite{RecPr}.

Thus, property (\ref{trk}) is valid for all kernels in the layer $ \lambda $. From the equivalence of (\ref{trk}) and (\ref{q1E1P4.11}), it follows that the temperature relations hold for all MEs with all indices $ l $, $ l_1 $, $ l_2 $.

\section{Conclusion}

As noted above, an essential difficulty
in the implementation of the moment method is to calculate the matrix elements of the collision integral.  In case of a strong deviation from the equilibrium 
we need to take into account the large number of terms in the expansion of DF in Barnett functions 
and, therefore, use MEs with larger indices. 
In our paper an algorithm for constructing the MEs for an arbitrary interaction potential between the particles with an arbitrary mass ratio is proposed with isotropic MEs assumed to be known. We emphasize that the non-linear collision integral is considered, which makes it possible to describe the evolution of strongly non-equilibrium systems. This can be important, for example, in describing the threshold reactions.
 In order to check developed procedure  we compared calculated MEs with MEs obtained before by other method
for the model of hard spheres \cite{book, ender2009}. The results were in a good agreement. 
Note that in the procedure described above only velocity recurrence relations obtained in \cite{book, ender2007} are used. They do not depend on the particles interaction law, being universal in this sense. Temperature recurrence relations for the constructed MEs are identities, the proof of which is  given in the last section of the article. 
The procedure for constructing isotropic MEs for any interaction potential  by the use of $ \Omega $ -integrals will be discussed in our next paper.

\section{Acknowledgments}
One of the authors (E.Yu.Flegontova) would like to express her gratitude to the Russian Fond of Basic Researches (RFBR) under Project No. 15-08-03440.
\section*{References}

\bibliography{mybibfile}

\end{document}